\documentclass{article}

\usepackage{graphicx}
\usepackage{color}
\usepackage{cite}
\usepackage{amssymb,amsfonts}
\usepackage{amsmath}
\usepackage{graphicx}
\usepackage{booktabs}
\usepackage{mathtools}
\usepackage{lineno}
\usepackage{float}
\usepackage{authblk}
\usepackage{caption,subcaption,setspace}
\usepackage{xspace}
\usepackage[bottom]{footmisc}
\usepackage[margin=1in]{geometry}

\def\naive{{na\"{\i}ve}\xspace}

\begin{document}

\title{A simple method for improving the accuracy of Chung-Lu random graph generation}

\author{Christopher Brissette, George M. Slota \thanks{Computer Science Department, Rensselaer Polytechnic Institute, Troy, NY
  (brissc@rpi.edu, slotag@rpi.edu).}}

\maketitle

\begin{abstract}
\noindent Random graph models play a central role in network analysis. The Chung-Lu model, which connects nodes $v_i,v_j$ based on their expected degrees $w_i$,$w_j$ according to the probability $\frac{w_iw_j}{\sum w_k}$ is of particular interest. It is widely used to generate null-graph models with expected degree sequences as well as implicitly define network measures such as modularity. Despite its popularity, practical methods for generating instances of Chung-Lu model-based graphs do relatively poor jobs in terms of accurately realizing many degree sequences. We introduce a simple method for improving the accuracy of Chung-Lu graph generation. Our method uses a Poisson approximation to define a linear system describing the expected degree sequence to be output from the model using standard generation techniques. We then use the inverse of this system to determine an appropriate input corresponding to the desired output. We give a closed form expression for this inverse and show that it may be used to drastically reduce error for many degree distributions. 
\end{abstract}

\section*{Introduction}
Say we wish to generate a random graph $G=(V,E)$ with a degree distribution $y=\{N_1,N_2,\cdots,N_m\}$ where $N_k$ represents the numbers of nodes with degree $k$. This is a problem that arises in many network science applications, most notably for the generation of null-models used for basic graph analytics~\cite{milo2002network}. Generating such networks exactly using e.g., a configuration model is computationally expensive for even moderately large networks. As such, we rely on probabilistic methods for large-scale graph generation that only match $y$ in expectation. The Chung-Lu random graph model \cite{chung2002average} is one such widely-used probabilistic model. This model pre-assigns to each node $v_i \in V(G)$ a weight $w_i$ corresponding the the degree we wish for the node to have and connects all nodes pairwise with the probability $p_{ij} = \frac{w_iw_j}{\sum w_k}$. There are a number of ways that generating such graphs can be done computationally. Some methods generate loops and multi-graphs, while others generate simple graphs. We focus on what is sometimes called the Bernoulli Method for generating Chung-Lu graphs~\cite{winlaw2015depth}, as it is amenable to the \emph{edge-skipping} technique~\cite{es_chunglu} that allows linear work complexity and near-constant parallel time for scalable implementations~\cite{alam2016efficient,slota_sc2019,garbus2020parallel}. In this method, we implicitly consider all $\frac{n(n-1)}{2}$ pairs of edges between unique nodes and generate edge $(v_i,v_j)$ with $i\neq j$ according to the probability $p_{ij}$. This generates a simple-graph with degree sequence $\Tilde{y}$ where $\mathbb{E}[\Tilde{y}] = y$.

\begin{figure}[!ht]
     \centering
     \includegraphics[width=0.42\textwidth]{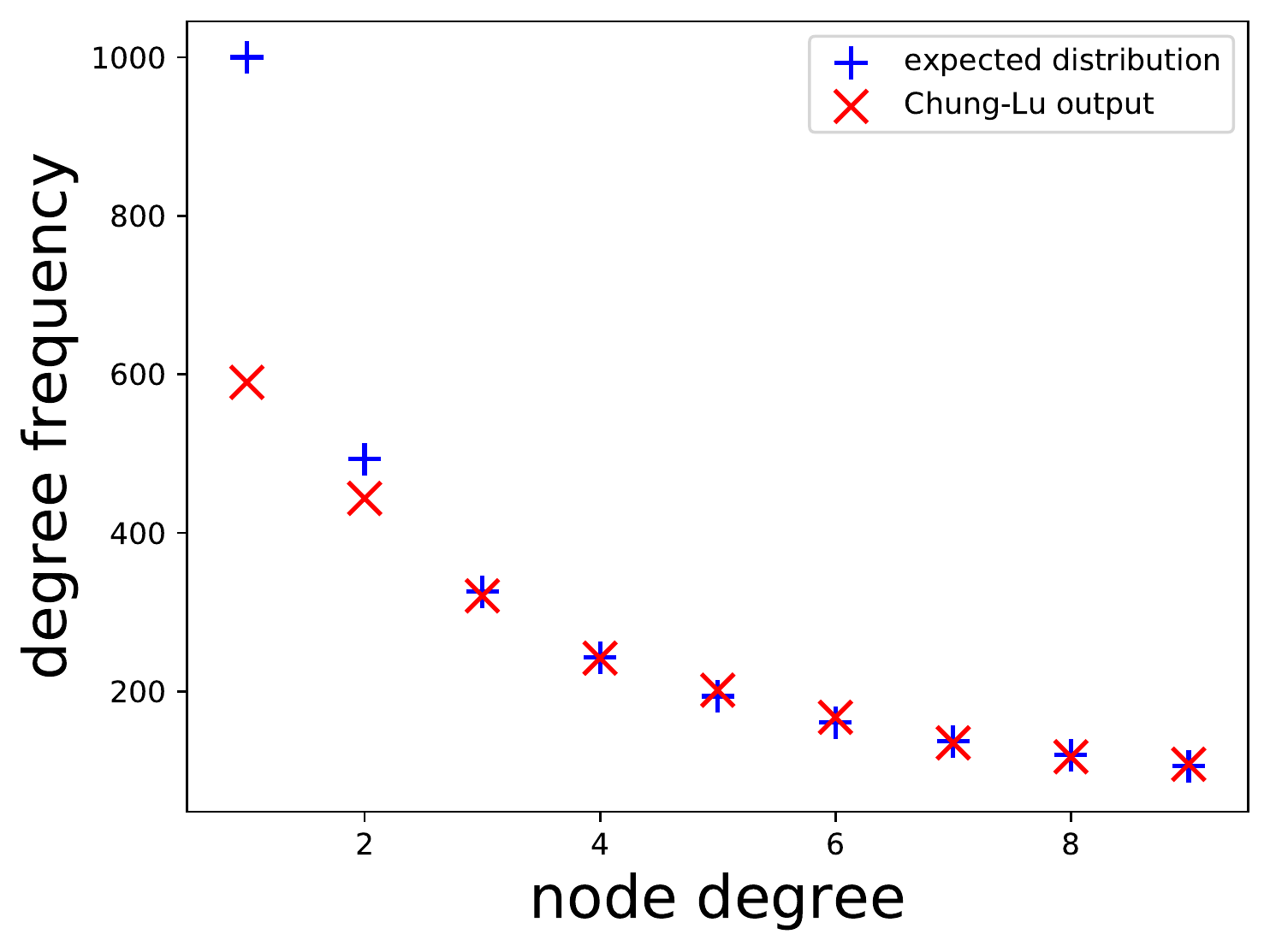}
     \includegraphics[width=0.42\textwidth]{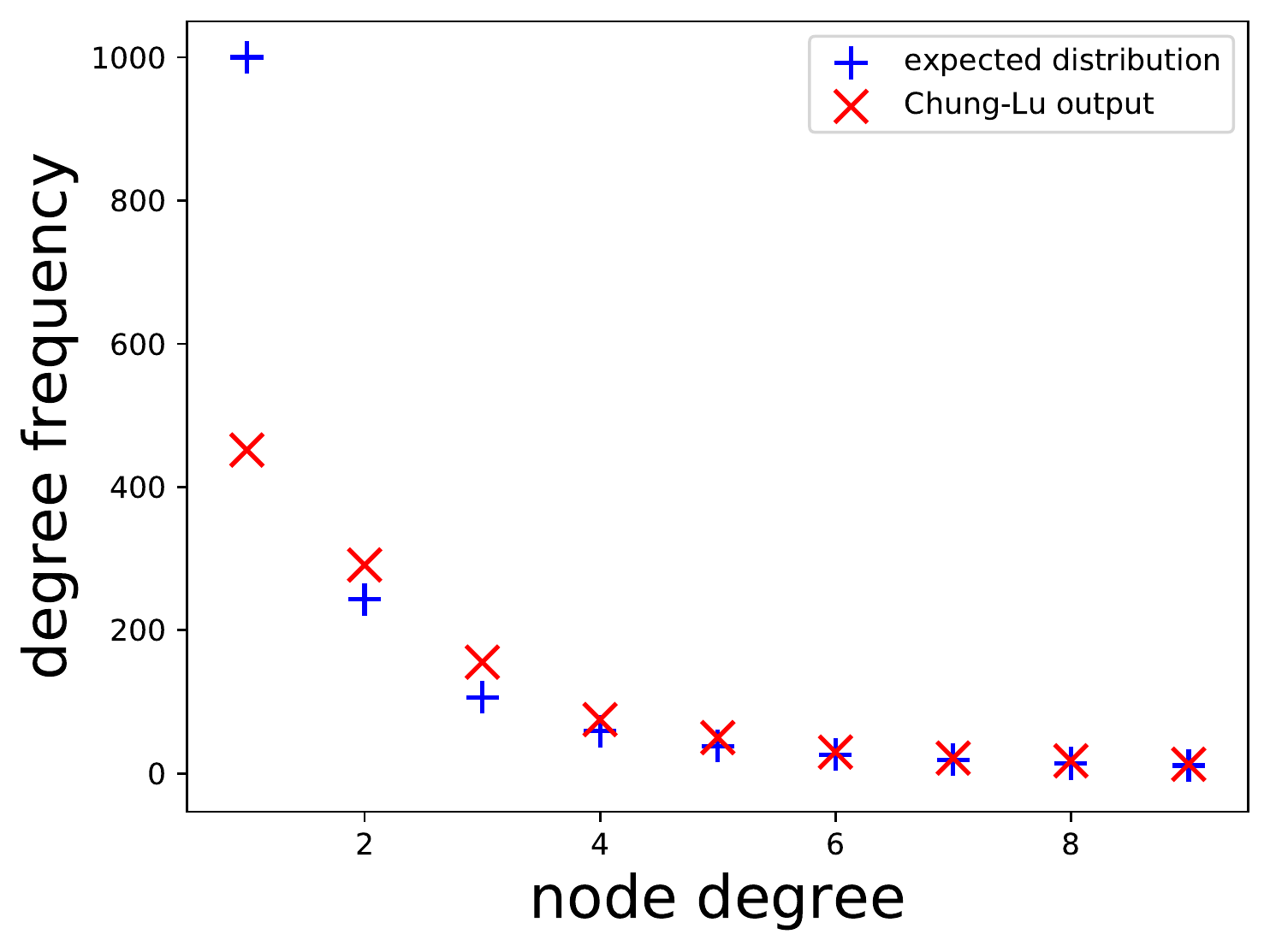}
     \includegraphics[width=0.42\textwidth]{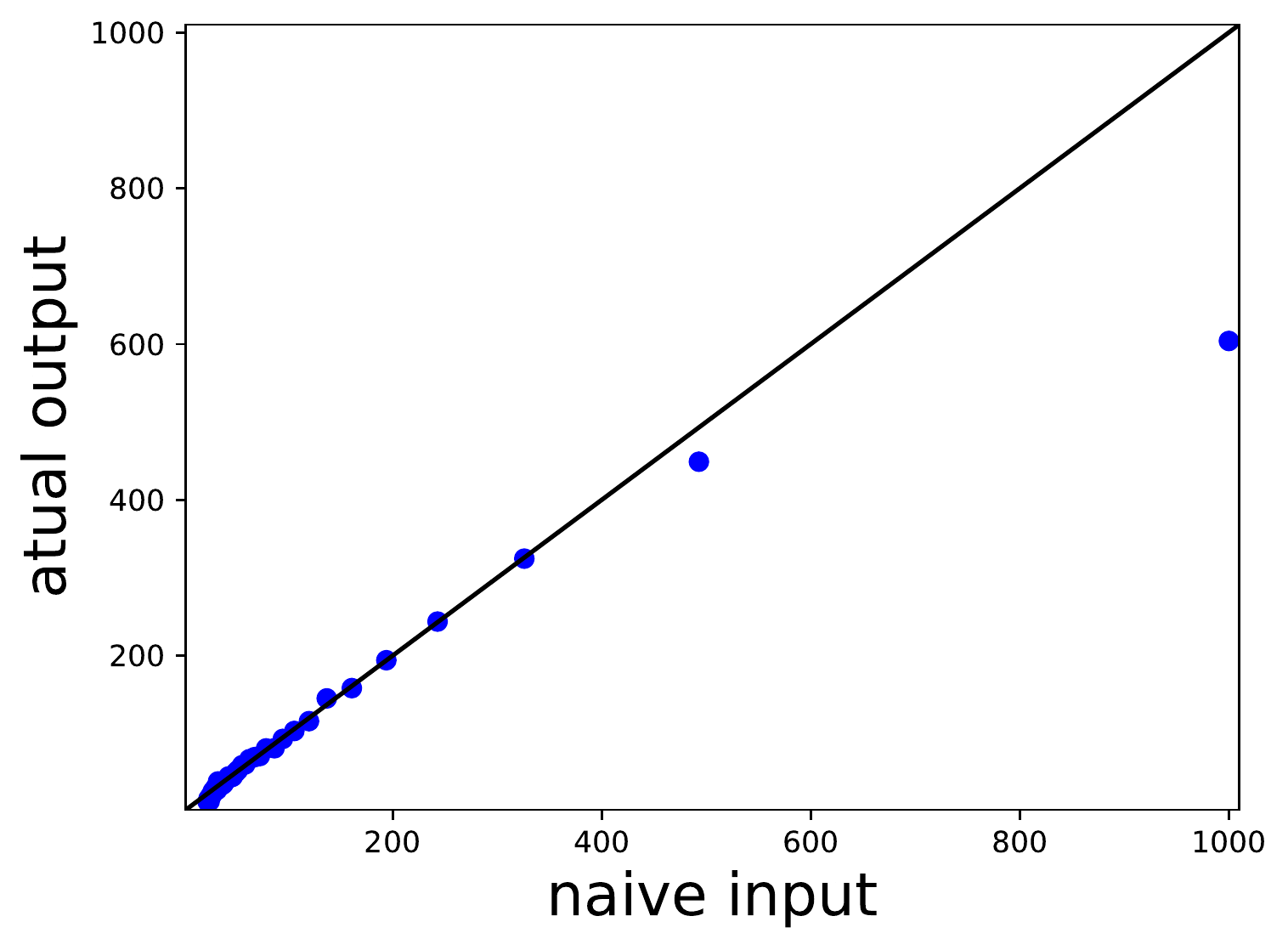}
     \includegraphics[width=0.42\textwidth]{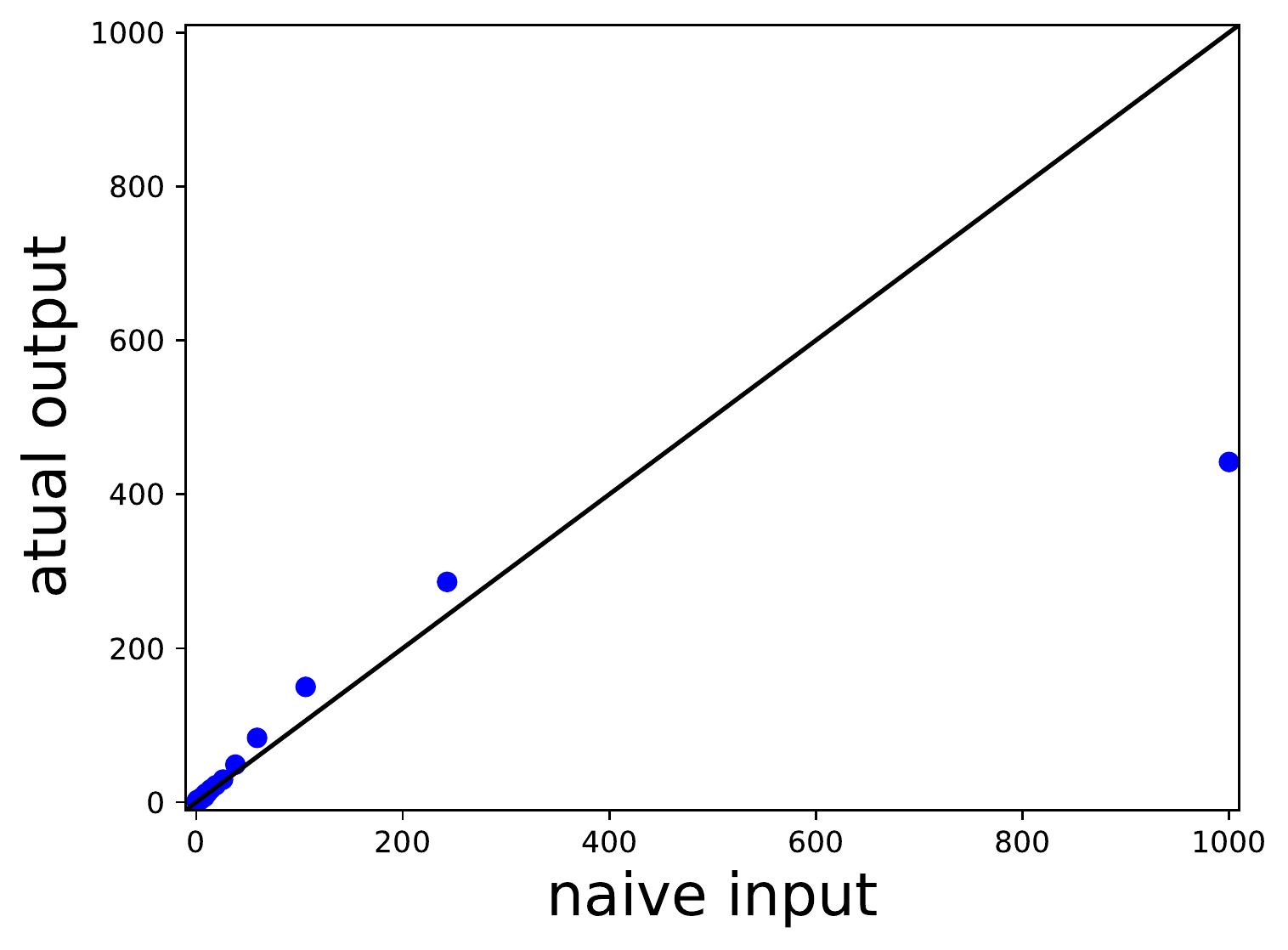}
     
     \caption{\textbf{Distribution error of Chung-Lu.} In the top plots, we consider the degree classes between one and nine for two different power law distributions. On the left is a power-law distribution with exponent $\beta = 1.0$ and on the right is a power-law distribution with exponent $\beta = 2.0$. In the top two plots, crosses represent the input distribution and x's represent the average distribution for 20 instances of Chung-Lu graphs given the input distribution distribution as input. We can see that the Chung-Lu generated graphs drastically under-represent degree one nodes. This is a phenomenon that commonly occurs in application and can greatly affect generation accuracy, given that power law degree distributions widely occur in nature and tend to have many low-degree nodes. The bottom two plots show how closely the Chung-Lu output distribution matches the input distribution up to the degree class 40. In both plots we see that most of the distribution is matched well, however a few degree classes cause almost all of the distribution error.}
     \label{fig:diffplots}
 \end{figure}

\indent The Chung-Lu model, though popular and theoretically sound under the tame condition that ${w_iw_j} < \sum_{k=1}^{m} w_k$ for all $v_i,v_j \in V$, can produce degree distributions drastically different from the desired expectation in practical settings, as is widely noted and addressed in the literature~\cite{winlaw2015depth,chodrow2020moments,garbus2020parallel,britton2006generating,van2013critical,pfeiffer2012fast,durak2013scalable}. This is particularly challenging when Chung-Lu generation is utilized as a subroutine for more complex graph generation, such as when generating graphs that also match a clustering coefficient distribution (e.g., the BTER model)~\cite{bter} or a community size distribution for community detection benchmarking~\cite{slota2020parallel,slota_sc2019}. See Figure~\ref{fig:diffplots} as an example of the observed error when generating some graphs. As can be seen, the output of Chung-Lu in both cases underestimates the number of degree one nodes, and accrues additional error from other low degree families as well. Distribution errors are not particularly surprising given that the model is inherently probabilistic, however they do pose potential issues. This suggests that instead of strictly caring about the expected degree of each node in Chung-Lu generation, as is generally done, we should additionally consider deeper statistical properties of the model in application.

\indent To get a better idea about the output of Chung-Lu, consider grouping all nodes by expected degree. That is, take degree families $d_k = \{v_i \in V : w_i = k\}$ and consider connections between them. From the point of view of a single node $v_j \in V$ with expected degree $w_j$ the number of connections it has to each degree family $d_k$ is bonomially distributed with mean $\frac{kw_j}{\sum w_i}|d_k|$. Therefore the degree distribution of each node in $d_{w_j}$ is the sum of $m$ independent binomial random processes where $m$ is the maximum expected degree of the graph. This provides additional statistical information about the degree distribution beyond the mean that may be used to improve the accuracy of Chung-Lu graph generation.

\indent Since the degree distribution of each degree family $d_k$ in our graph is binomially distributed, we may apply a further approximation. Because the limiting case of the binomial distribution is the Poisson distribution, we approximate the number of connections between nodes in a given degree family with all other nodes as a sum of Poisson distributions, which is again Poisson. We note that often times a desired degree distribution $D$ will be such that certain degree classes will not have the number of nodes required for this approximation to have guaranteed accuracy. In fact, power-law degree distributions will in general have degree families $d_k$ where $k \approx m$ such that $|d_k| \approx 1$. However, we also note that this additionally means the node-wise error contributed by those families is relatively small, so we are willing to sacrifice some accuracy in lieu of a cleaner description.

\indent Say that $X_{ij}$ is the Poisson distribution representing the degree of each node in $d_i$ if $d_i$ only connected to nodes in $d_j$. Additionally, take the mean of this Poisson distribution to be $\gamma_{ij}$. Because the means of independent Poisson distributions are additive, we have the following linear system, describing the means of each distribution.
\begin{equation}
    \begin{bmatrix}
    \gamma_{11} & \gamma_{12} & \cdots & \gamma_{1m} \\
    \gamma_{21} & \gamma_{22} & \cdots & \gamma_{2m} \\
    \vdots      & \vdots      & \ddots & \vdots      \\
    \gamma_{m1} & \gamma_{m2} & \cdots & \gamma_{mm}
    \end{bmatrix} 
    \begin{bmatrix}
    1\\ 
    1\\
    \vdots \\
    1
    \end{bmatrix}
    =
    \begin{bmatrix}
    \mu_1\\
    \mu_2\\
    \vdots \\
    \mu_m
    \end{bmatrix}
    \label{eq:means}
\end{equation}
This matrix provides additional rationale for our Poisson approximation. Since we assumed the distributions were Poisson we may now add means of Poisson distributions directly as opposed to computing with more complex independent binomial distributions. In the case of the Chung-Lu model, each $\gamma_{ij} = \frac{w_iw_j}{\sum w_k}$. This, perhaps as expected, gives the right hand means of $\mu_k = k$. This means that the degrees within each degree family $d_k$ should be approximately Poisson distributed. Before moving on we note that a similar analysis can be done for any connection probabilities. While we are focusing on Chung-Lu probabilities, this model also describes the output degree sequence for any set of chosen $p_{ij}$ between degree families, albeit with potential changes to the means $\mu_i$.

\indent Given the description offered by Equation~\ref{eq:means} we now have the tools to estimate the output of Chung-Lu through a simple linear system. Consider an input degree distribution $y=[N_1,N_2,\cdots,N_m]^T$ as a vector in $\mathbb{R}^m$ with $N=\sum_{i=1}^mN_i$. Additionally take $poiss(k)$ to be the probability density function of the Poisson distribution with mean $k$. We can calculate the expected output $\Tilde{y}$ of this as follows.
\begin{equation}
    \textbf{Q}y = \begin{bmatrix}
    | & | &   & | \\
    poiss(1) & poiss(2) & \cdots & poiss(m) \\
    | & | &    & |
    \end{bmatrix}
    \begin{bmatrix}
    N_1 \\
    N_2\\
    \vdots \\
    N_m
    \end{bmatrix} = \Tilde{y}
    \label{eq:full_poiss}
\end{equation}
Note here we are assuming $poiss(k)$ is the full, discrete version of the Poisson distribution with mean $k$. This implies that the system in Equation~\ref{eq:full_poiss} maps $\mathbb{R}^m$ to an infinite dimensional space. This is obviously not ideal, since we wish to use this matrix in estimating Chung-Lu outputs computationally. We therefore truncate the Poisson matrix $\textbf{Q}$ to be square in $\mathbb{R}^{m\times{m}}$ by removing the first row corresponding to degree zero nodes, as well as everything below the $m^{th}$ row. We will denote this matrix by $\textbf{P}$. Our justification for this truncation is two-fold. One, we are inputting a degree distribution in $\mathbb{R}^m$, and we mainly only care about error with regards to those output degrees between one and $m$. Two, making the matrix square allows for us to invert the matrix which will be useful for generating Chung-Lu graphs with more accurate degree sequences. Note that truncating $\textbf{Q}$ to some dimension $m$ amounts to ignoring nodes with degree zero as well as nodes with degree higher than $m$. If we wish to obtain error information for higher degrees as well we can easily append zeros to the end of our input distribution and consider $\textbf{P}{\in}\mathbb{R}^{n\times{n}}$ where $n > m$ and $m$ is the maximum degree of our desired distribution. Then, for large enough $n$, our error is only ignoring nodes of degree zero. In a practical setting, these nodes would be thrown out and ignored, anyways. The rest of this paper discusses properties of $\textbf{P}$ and how it can be used to improve the accuracy of Chung-Lu outputs.

\subsection*{Our Contributions}
As noted, while Chung-Lu graph generation is a useful tool for many theoretical purposes and is used widely in fields such as social network analysis, it often does a poor job of approximating distributions at the ends. The specific issue considered in this paper is that Chung-Lu generated networks will often under-represent low degree nodes. In Figure~\ref{fig:diffplots}, we can clearly see that actual Chung-Lu realizations may easily contain less than 60 percent of the desired number of degree one nodes. This can lead to a great deal of inaccuracy for distributions with particularly large numbers of low degree nodes. In practice this generally means that generated graphs will have many vertices of degree zero, so one way of resolving this issue is to connect these nodes to the graph in order to inflate the number of degree one nodes. Depending on the degree distribution this can easily skew other degree classes without careful choice of where these nodes are connected. This may require considerable computation. For this reason, it is far simpler in application to throw away degree zero nodes.

\indent For this reason we suggest the matrix model referenced in the introduction. The standard input distribution for Chung-Lu is simply the desired output distribution $y$. We suggest a ``shifted'' Chung-Lu algorithm where, given a matrix model $\textbf{P}$ for the output of the Chung-Lu algorithm, we take our desired output distribution $y$ and solve for $x = \textbf{P}^{-1}y$. Then the input to a Chung-Lu graph generator is $x$ as opposed to the desired output. This is particularly compelling since the matrix $\textbf{P}^{-1}$ only depends on the maximum degree of our desired output distribution and once computed allows for drastic accuracy improvement at negligible algorithmic cost.

\indent In addition, we prove several notable properties about the baseline model and our approaches when used in a generative setting. Most interestingly, we note that the \naive model is actually quite limited in the scope of graphs that it can generate reliably. 

\section*{Properties of the matrix model}
\subsection*{A factorization of P}
From the introduction, we use the assumption that the degree distribution of each family is approximately Poisson distributed to form a matrix that will transform input distributions into approximate output distributions from the Chung-Lu model. Assume that our input distribution has degrees in $\mathbb{N}_m=\{1,\cdots,m\}$ and is represented by $x = [N_1,N_2,\cdots,N_m]^T$ where $N_k$ represents the number of nodes with expected degree $k$ and $N=\sum_{k=1}^mN_k$. Using $poiss(k,\cdot)$ to represent the discrete Poisson distribution with mean $k$, we represent our matrix \textbf{P} as follows.  
\begin{align}
\begin{split}
    \textbf{P} &= 
    \begin{bmatrix}
    poiss(1,1) & poiss(2,1) & poiss(3,1) & \cdots & poiss(m,1) \\
    poiss(1,2) & poiss(2,2) & poiss(3,2) & \cdots & poiss(m,2) \\
    \vdots & \vdots & \vdots & \ddots & \vdots \\
    poiss(1,m) & poiss(2,m) & poiss(3,m) & \cdots & poiss(m,m) 
    \end{bmatrix} \\
    &= 
    \begin{bmatrix}
    e^{-1} & 2e^{-2} & 3e^{-3} & \cdots & me^{-m} \\
    \frac{e^{-1}}{2} & \frac{2^2e^{-2}}{2} & \frac{3^2e^{-3}}{2} & \cdots & \frac{m^2e^{-m}}{2} \\
    \vdots & \vdots & \vdots & \ddots & \vdots \\
    \frac{e^{-1}}{m!} & \frac{2^me^{-2}}{m!} & \frac{3^me^{-3}}{m!} & \cdots & \frac{m^me^{-m}}{m!} 
    \end{bmatrix} \\
    &= 
    \begin{bmatrix}
    1 & 0 & 0 & \cdots & 0 \\
    0 & \frac{1}{2!} & 0 & \cdots & 0 \\
    \vdots & \vdots & \vdots & \ddots & \vdots \\
    0 & 0 & 0 & \cdots & \frac{1}{m!}
    \end{bmatrix}
    \begin{bmatrix}
    1 & 1 & 1 & \cdots & 1 \\
    1 & 2 & 3 & \cdots & m \\
    \vdots & \vdots & \vdots & \ddots & \vdots \\
    1 & 2^{m-1} & 3^{m-1} & \cdots & m^{m-1}
    \end{bmatrix}
    \begin{bmatrix}
    e^{-1} & 0 & 0 & \cdots & 0 \\
    0 & 2e^{-2} & 0 & \cdots & 0 \\
    \vdots & \vdots & \vdots & \ddots & \vdots \\
    0 & 0 & 0 & \cdots & me^{-m}
    \end{bmatrix} \\ 
    &= \textbf{A}\textbf{V}\textbf{B} \\
\end{split}
\label{eq:P_fact}
\end{align}
Note that realizing a random Chung-Lu graph model amounts to computing $\textbf{P}x$ for some pre-defined $x$. We instead look at the inverse problem of determining $x{\in}\mathbb{R}^{+m}$ given $\textbf{P}{\in}\mathbb{R}^{m\times{m}}$ and desired output $y{\in}\mathbb{R}^{+m}$. Here $\mathbb{R}^{+m}$ is the positive region of $\textbf{R}^m$. This amounts to solving the linear system $\textbf{P}x = y$. One may be tempted to simply invert this matrix using any number of computational methods, and this is reasonable for small $m$. However, given the factorization in Equation~$\ref{eq:P_fact}$, we have that $\textbf{P} = \textbf{A}\textbf{V}\textbf{B}$ with $\textbf{V}$ a Vandermonde matrix. Due to the extremely poor conditioning of both $\textbf{A}$ and $\textbf{V}$, using a computational method for inverting $\textbf{P}$ is not advised. Fortunately $\textbf{A}$ and $\textbf{B}$ are diagonal, meaning they are easy to invert, so finding the inverse of $\textbf{P}$ only requires finding an inverse to $\textbf{V}$. That is, we would like to find the following.
\begin{align}
    \textbf{P}^{-1} &= 
    \begin{bmatrix}
    \frac{1}{e^{-1}} & 0 & 0 & \cdots & 0 \\
    0 & \frac{1}{2e^{-2}} & 0 & \cdots & 0 \\
    \vdots & \vdots & \vdots & \ddots & \vdots \\
    0 & 0 & 0 & \cdots & \frac{1}{ne^{-n}}
   \end{bmatrix}
    \textbf{V}^{-1}
    \begin{bmatrix}
    1 & 0 & 0 & \cdots & 0 \\
    0 & 2! & 0 & \cdots & 0 \\
    \vdots & \vdots & \vdots & \ddots & \vdots \\
    0 & 0 & 0 & \cdots & n!
    \end{bmatrix}
    \label{eq:inv_fact}
\end{align}
Again, we do not want to compute this using standard computational methods, since Vandermonde matrices are the textbook examples of nearly uninvertible matrices. Fortunately, our Vandermonde matrix is such that it has a special structure yielding a somewhat simple closed-form inverse given in \cite{eisenberg98}. It relates each entry in the matrix to associated binomial coefficients and Stirling numbers of the first kind. Explicitly, each entry is expressed as follows.
\begin{equation*}
    \textbf{V}_{ij}^{-1} = (-1)^{i+j}\overset{n}{\underset{k=max(i,j)}{\sum}}\frac{1}{(k-1)!}{k-1 \choose i-1}\genfrac{[}{]}{0pt}{}{k}{j}
\end{equation*}
We can now compute the input of Chung-Lu that will best approximate the desired output according to $\textbf{P}^{-1}y = x$. While this is powerful in its own right, caveats still remain.

\subsection*{Integer solutions do not exist}
Recall that $x$ and $y$ refer to our input degree distribution and output degree distribution respectively. Ideally $x$ and $y$ will have only positive integer entries corresponding to numbers of nodes, otherwise we will be forced to do some rounding. Consider the system $\textbf{P}x = y$. We rewrite this as $\textbf{B}^{-1}x = \textbf{V}^{-1}\textbf{A}^{-1}y$ and generalize the left hand side to the form in Equation~\ref{eq:system}. Here $t_k$ is an arbitrary integer. If this system does not have integer solutions $\{y_1,y_2,\cdots,y_m\}\in{\mathbb{N}}$, then the stricter linear system given by $\textbf{P}x = y$ certainly will not.
\begin{equation}
    \begin{bmatrix}
    t_1e^{-1} \\
    t_2e^{-2} \\
    \vdots \\
    t_ne^{-m}
    \end{bmatrix}
    =
    \textbf{V}^{-1}
    \begin{bmatrix}
    y_1 \\
    2!y_2 \\
    \vdots \\
    m!y_m
    \end{bmatrix}
    \label{eq:system}
\end{equation}
We examine individual sums in the linear system. Assume $y_j$ and $t_k$ are positive integers. Then we wish to find solutions to the following equation.
\begin{align}
\begin{split}
    t_ke^{-k} &= \overset{m}{\underset{j=1}{\sum}}j!\textbf{V}_{kj}^{-1}y_j \\
    \Rightarrow e^{-k} &= \overset{m}{\underset{j=1}{\sum}}\frac{j!\textbf{V}_{kj}^{-1}y_j}{t_k}
\end{split}
\label{eq:not_integer}
\end{align}
By inspection we see that the right hand expressions $\frac{j!\textbf{V}_{kj}^{-1}y_j}{t_k}$ cannot be irrational. This is because $y_j$ and $t_k$ are positive integers by assumption. Therefore this means that for this system to have solutions requires $e^{-k}$ to be equal to a finite sum of rational numbers for $k\in{\mathbb{Z}}$. However it is known that for such exponents $e^{-k}$ is irrational. Therefore there are no exact integer solutions to this linear system. We do note however that given a desired output $y$ we can find an input $x$ ``close'' to an integer value with negligible algorithmic cost, we discuss how ``close'' this solution is later in the paper.
\subsection*{Not all solutions are positive}
\begin{figure}[!ht]
     \centering
     \includegraphics[width=0.81\textwidth]{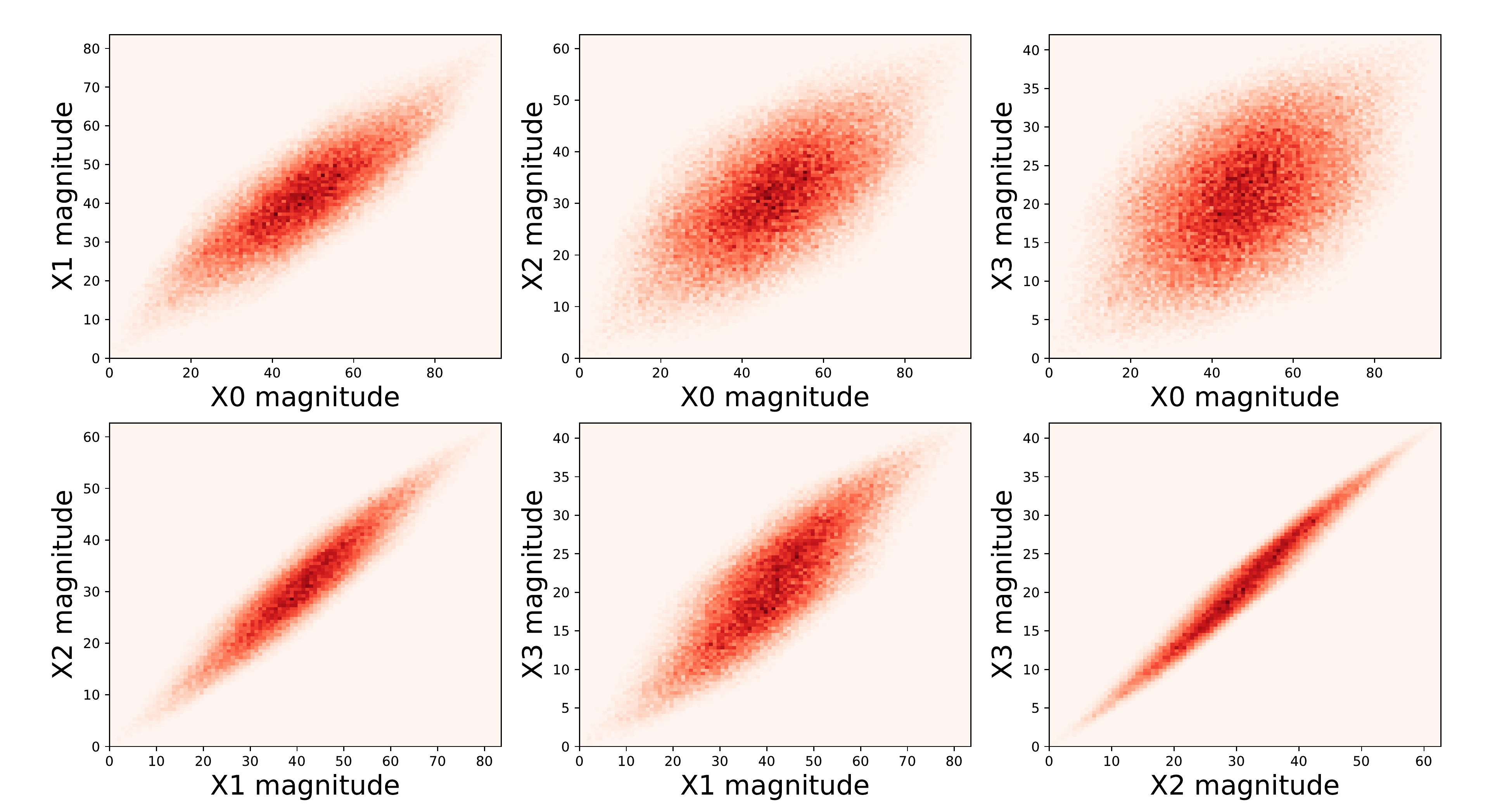}
     \caption{\textbf{Action of $\textbf{P}$ on the positive hypercube.} Here we can see plots of projections of random vectors under the action of $\textbf{P}$ as a heat map. The sample consists of 100,000 random vectors with random integer entries selected to be within $\{0,\cdots,100\}$ under the action of $\textbf{P}\in{\mathbb{R}^{4\times{4}}}$. The output vectors are then projected onto each canonical unit vector $e_j\in{\mathbb{R}^4}$ and plotted pairwise. These vectors are referred to as $Xi$ in the axis labels. Intuitively this shows all feasible output from a Poisson random graph model with node degrees limited to those in $\{1,2,3,4\}$. We can see that all positive vectors remain inside the positive region as expected, and we also see how sharply limiting this is for finding positive solutions of $\textbf{P}^{-1}y$ for $y$ positive.}
     \label{fig:2dproj}
 \end{figure}

 $\textbf{P}$ has only positive real entries. This implies that for any element-wise positive vector $x$, $\textbf{P}x$ is also positive. While this implies that any positive input will yield an approximately valid result, it does not exclude the possibility of vectors with negative entries also mapping into the positive region of $\mathbb{R}^{m}$ under the action of $\textbf{P}$. This means that we may not be able to use the output of $\textbf{P}^{-1}y=x$ as the input of $\textbf{P}x$ since $x$ has the possibility of containing negative elements. In Figure~\ref{fig:2dproj}, we can see what the action of $\textbf{P}$ looks like on a sample of random vectors for $\textbf{P}{\in}\mathbb{R}^4$. Notice how, as expected, it ``squishes'' the positive region into a small sliver.
 \indent Under what circumstances will the solutions of $\textbf{P}^{-1}y$ be positive for element-wise positive input $y$? This question is akin to asking what vectors compose the dark red regions in Figure~\ref{fig:2dproj}. We can classify these regions by their mean. This is a straightforward computation that consists of integrating $\frac{1}{r^m}\textbf{P}x$ over all the vectors in a positive cube in $\mathbb{R}^m$ with volume $r^m$. Call $\mu_m(r)$ the mean action of $\textbf{P}$ on the positive hypercube with volume $r^m$, then we have the following example in $\mathbb{R}^2$.
 \begin{align}
 \begin{split}
     \mu_2(r) &= \frac{1}{r^2}\int_{0}^{r}\int_{0}^r \textbf{P}[x_1,x_2]^T \,dx_1 dx_2 \\
     &=
     \frac{1}{r^2}\int_{0}^{r}\int_{0}^r 
     \begin{bmatrix}
     poiss(1,1) & poiss(2,1) \\
     poiss(1,2) & poiss(2,2)
     \end{bmatrix}
     [x_1,x_2]^T \,dx_1 dx_2 \\
     &=
     \frac{r}{2}[poiss(1,1)+poiss(2,1),poiss(1,2)+poiss(2,2)]^T \\
     &= \frac{r}{2}\textbf{P}\textbf{1} 
     \end{split}
     \label{eq:mean} 
 \end{align}
 Here, $\textbf{1}$ is the vector in $\mathbb{R}^m$ of all ones. This is in fact the general expression for the mean action of $\textbf{P}$ on the positive hypercube with sides of length $r$ regardless of dimension. That is, $\mu_m(r)=\frac{r}{2}\textbf{P}\textbf{1}$ for any $m{\in}\mathbb{N}$ and $\textbf{1}{\in}\mathbb{R}^m$. While the mean $\mu_m(r)$ is such that $\textbf{P}^{-1}\mu_m(r)$ is a member of the positive reals in $\mathbb{R}^m$, this defines a very small class of possible graphs, and does not include every feasible network. 
 
 \indent Instead given a number of nodes $N$ we look to bound how many nodes of each degree are feasible. That is, if we have some degree distribution $x$ with L1-norm $\|x\|_1 = N$ we wish to find lower and upper bounds, $l_i$ and $u_i$ respectively on $|(\textbf{P}x)_i|$ such that $l_i \leq |(\textbf{P}x)_i| \leq u_i$. We want to do this for every degree family. Take the projector $\rho_i = e_i^Te_i$ where $e_i$ is the $i^{th}$ canonical unit vector in $\mathbb{R}^m$. Then we know $|(\textbf{P}x)_i| = \|\rho_i\textbf{P}x\|_1$. This directly implies from the structure of $\textbf{P}$ that for $\textbf{P}_{ij} = \underset{k}{min}|P_{ik}|$ we have,
 \begin{equation}
     N\textbf{P}_{ij} \leq \|\rho_i\textbf{P}x\|_1 \leq N\textbf{P}_{ii} \text{ for all } x{\in}\mathbb{R}^m \text{ such that } \|x\|_1 = N
     \label{eq:bounds}
 \end{equation} \\
 \indent While the bounds in Equation~\ref{eq:bounds} are useful to heuristically understand what kinds of distributions we can feasibly generate, the bounds are not tight. However, we can tighten them. Take the set of $m$ points, $E_N = \{Ne_k : e_k \text{ a canonical unit vector in } \mathbb{R}^{+m} \}$. We will call the hyperplane containing $E_N$ by the name $H_N$. The graphs we can possibly generate are defined by the intersection of the hyperplane $\textbf{P}H_N$ containing $\textbf{P}E_N$ and the convex shape defined by $R_N = \{v{\in}\mathbb{R}^{+m} : \textbf{P}_{kj_k} \leq v_k \leq \textbf{P}_{kk} \text{ where } \|v\|_1 = N \text{ and } \textbf{P}_{kj_k} = \underset{t}{min}|P_{kt}|\}$. That is, the graphs we can generate exist within,
 \begin{equation}
     R_N \cap \textbf{P}H_N
     \label{eq:intersection}
 \end{equation}
 
 Therefore given a desired output degree distribution $y$ we can check if it has a meaningful, positive inverse $x = \textbf{P}y {\in} \mathbb{R}^{+m}$ by checking if it is within the intersection defined in Equation~\ref{eq:intersection}. Fortunately $\textbf{P}H_N$ is easy to solve for numerically, consider the following matrices,
 \begin{align}
     \textbf{B}_N &= 
     \frac{N}{m}
     \begin{bmatrix}
      | & | &  & | \\
      \textbf{1} & \textbf{1} & \cdots & \textbf{1} \\
      | & | &  & | 
     \end{bmatrix} \in \mathbb{R}^{m\times{m}} \\
     \Delta_N &= (N\textbf{P} - \textbf{P}\textbf{B}_N)^T
     \label{eq:delta} \\
      &= \left(N\textbf{I}-\frac{N}{m}\textbf{B}_N\right)\textbf{P}^T
 \end{align}
 Eigenvectors associated with zero eigenvalues of this matrix $\Delta_N\omega=0$ are then normal to our plane. We can see immediately that $\textbf{P}^T\omega=\textbf{1}$ uniquely defines the vector we are looking for. Therefore for a given desired output distribution $y$ we can determine if there is a valid-positive input distribution $x$ such that $\textbf{P}x = y$ by checking the condition in Equation~\ref{eq:bounds} and the following second condition where $(\cdot,\cdot)$ is the inner product, 
 \begin{equation}
     (y-\frac{N}{n}\textbf{P}\textbf{1},\omega) = 0 
     \label{eq:dotprod}
 \end{equation}
 We note that $\omega$ needs to be computed for $\textbf{P}{\in}\mathbb{R}^{m\times{m}}$. Luckily, for this method we are explicitly calculating $\textbf{P}^{-1}$ so we can simply compute $(\textbf{P}^{-1})^T\textbf{1}=\omega$ directly. We additionally note that we do not know apriori what value of $N$ is required to generate graph $y$, so theoretically $N$ has to be checked over all possible $N{\in}\mathbb{R}$. We can limit this space considerably by noting that $\|\rho_m\textbf{P}x\|_1 \geq N\textbf{P}_{mj}$ meaning $y_m \geq N\textbf{P}_{mj}$. Therefore we know $N$ is such that $\|y\|_1 \leq N \leq \frac{y_m}{\textbf{P}_{mj}}$ where $\textbf{P}_{mj}$ is as in condition~\ref{eq:bounds}. This drastically reduces our search space for condition~\ref{eq:dotprod} in the case of small $m$, but for large $m$, $\textbf{P}_{mj}$ is near zero meaning that effectively we only have a lower bound on what $N$ may be. In general, checking conditions~\ref{eq:bounds} and~\ref{eq:dotprod} are useful for determining if a graph $y$ may be generated, but it will likely be more reasonable in application to simply apply $\textbf{P}^{-1}$ to the distribution and see if the resulting vector is positive. 

\subsection*{Rounded solutions are local}
If we take  $Int(\cdot)$ to represent the nearest element-wise integer rounding of a vector, we have that $\|\textbf{P}^{-1}y - Int(\textbf{P}^{-1}y)\|_1 \leq \frac{1}{2}m$ where we know $\textbf{P}{\in}\mathbb{R}^{m\times{m}}$. For a network with a number of vertices $N \gg m$ this will likely be a negligible quantity, however an important thing to note is that this is the backwards problem. To actually generate networks, we instead consider the original system, $\textbf{P}x=y$; however, now we have an approximate rounded solution for $x$. This gives us the following where $\rho(\cdot)$ is the spectral radius and $\|\cdot\|$ is an arbitrary induced norm.
\begin{align*}
    \textbf{P}Int(\textbf{P}^{-1}y) &= \Tilde{y} \\
    Int(\textbf{P}^{-1}y) &= (x-\delta{x}) \\
    \Rightarrow \textbf{P}(x-\delta{x}) &= \Tilde{y} \\
    \Rightarrow \|y-\Tilde{y}\|  &= \|\textbf{P}\delta{x}\| \\
    &\leq \rho(\textbf{P})\|\delta{x}\| \\
    \rho(\textbf{P}) &\leq \|\textbf{P}\| \\
    &\leq 1 \tag{guaranteed by taking $\|\cdot\| = \|\cdot\|_{1}$} \\
    \therefore \|y-\Tilde{y}\| &\leq \|\delta{x}\| 
\end{align*}
Worded differently, this says that, given a desired output $y$, we call $\Tilde{x} = Int(\textbf{P}^{-1}y)$, ${x} = \textbf{P}^{-1}y$, and $\Tilde{y} = \textbf{P}\Tilde{x}$. Then, for $\epsilon > 0$ we have the following.
\begin{align*}
    \|x-\Tilde{x}\| = \|\textbf{P}^{-1}y - Int(\textbf{P}^{-1}y)\| &< \epsilon \\
    \Rightarrow \|y-\Tilde{y}\| &< \epsilon
\end{align*}
Meaning that, ignoring floating point arithmetic error, there is no additional error blow up present in the problem outside of that incurred by rounding the result of the backward problem. Therefore, we know in the worst case that,
\begin{equation*}
    \|y-\Tilde{y}\|_1 < \frac{1}{2}m.
\end{equation*}
The severity of this potential error is dependent on the value of $\|y\|_1$. Since we are concerned mainly with large and sparse networks in practical applications, we assume $m << \|y\|_1$ in general, so this error bound poses minimal accuracy concerns. However, we do note that in the dense case, the number of nodes $N$ in our graph is close to $m$ and this error bound poses significant issues. However, this should come as no surprise and can be viewed as a representation of us breaking the condition for Chung-Lu that $w_iw_j < \sum_{k=1}^m w_k$. This is unfortunately a byproduct of Chung-Lu-like random graph models and other methods have to be incorporated~\cite{chodrow2020moments} to generate dense random graphs. 

\subsection*{Numerical considerations}

Recall the form of $\textbf{V}_{ij}^{-1}$.
\begin{equation*}
    \textbf{V}_{ij}^{-1} = (-1)^{i+j}\overset{n}{\underset{k=max(i,j)}{\sum}}\frac{1}{(k-1)!}{k-1 \choose i-1}\genfrac{[}{]}{0pt}{}{k}{j}
\end{equation*}
Both ${k-1 \choose i-1}$ and $\genfrac{[}{]}{0pt}{}{k}{j}$ represent the binomial coefficient and Stirling number of the first kind, respectively. These are combinatorially large, which will pose issues for expressing this matrix in floating points as $n$ scales. Compounding the issue, the largest elements in $\textbf{A}^{-1}$ and $\textbf{B}^{-1}$ are on the orders of $n!$ and $e^{n}$ respectively, meaning that computing $\textbf{P}^{-1}$ in standard \textit{IEEE} floating point arithmetic is intractable for even a comparatively minuscule $m$.

\indent One way to deal with this issue is to artificially inflate the precision used in computation. This is the method we chose to use for our testing. We use \textsc{matlab}'s \texttt{vpa}$(\cdot)$ functionality within the symbolic toolbox. This allows us to inflate our precision as needed, however the precision needs to be increased significantly as our maximum degree $m$ increases. This should not matter too much however, since $\textbf{P}^{-1}$ is the same for all $m$, and can be computed once and stored as a symbolic matrix for later use. We recommend storing the individual matrices $\textbf{A}^{-1}$, $\textbf{B}^{-1}$, $\textbf{V}^{-1}$ from equation \ref{eq:inv_fact} instead of storing $\textbf{P}^{-1}$ since during testing we found that obtaining accurate outputs from $\textbf{P}^{-1}$ required greater precision than with it's factorization.

\section*{Results}
Our results focus on the predictive power of our matrix model $\textbf{P}$ as well as the error reduction due to using our ``shifted Chung-Lu'' input $\textbf{P}^{-1}y=x$ opposed to the \naive input $y$. We use the \textsc{python} package \texttt{networkx} and \texttt{networkx.expected\_degree\_graph($\cdot$)} to generate Chung-Lu random graphs, and \textsc{matlab} with \texttt{vpa($\cdot$)} to solve for our matrix model $\textbf{P}^{-1}y = x$. We specifically look at the proportional L1 error of \naive Chung-Lu generation and compare it with the proportional L1 error of shifted Chung-Lu generation.

\begin{figure}[!htb]
    \centering
    \includegraphics[width=\textwidth]{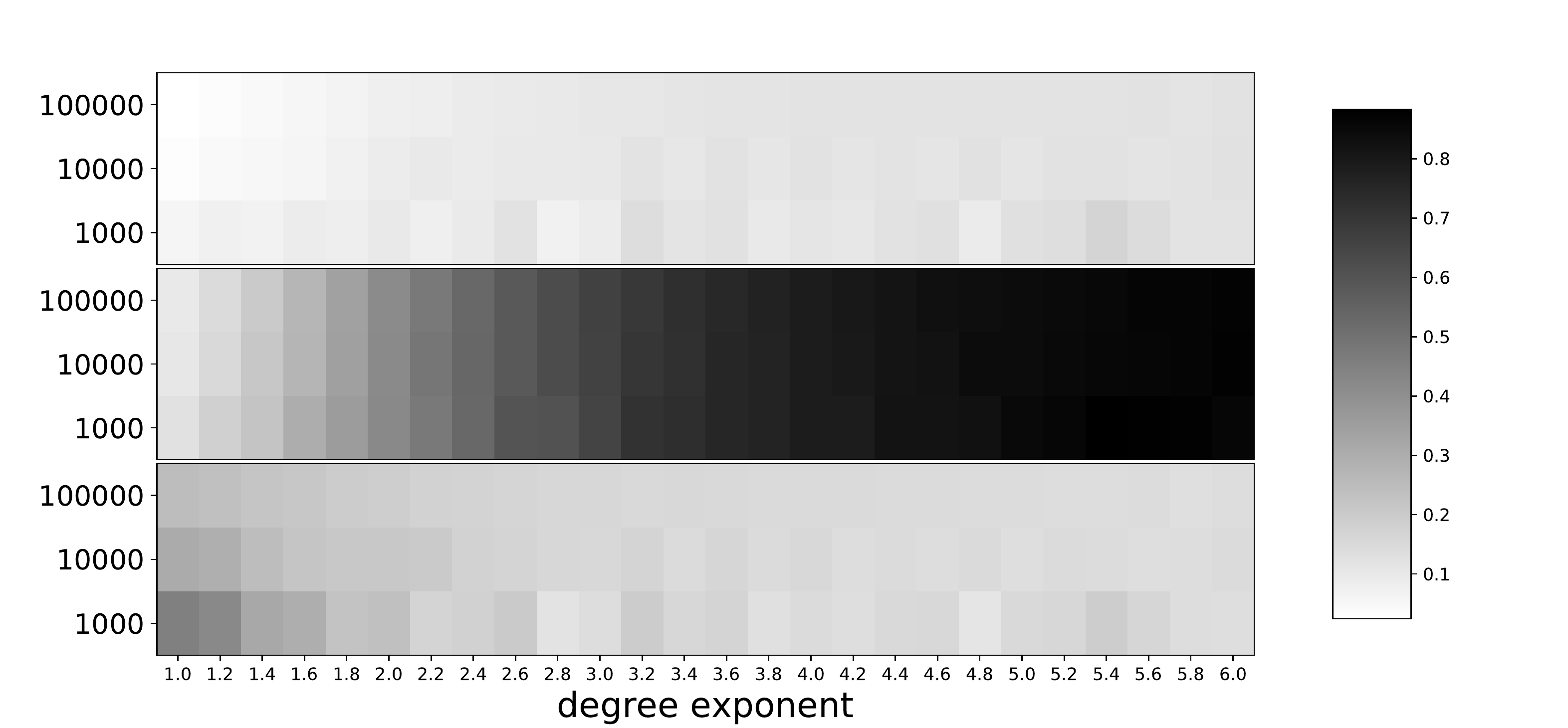}
    \caption{\textbf{Heatmaps of all involved errors.} Here we consider degree sequences of the form $Nk^{-\beta}$ where $\beta$ are along the x-axis and values of $N$ are along the y-axis. The Chung-Lu realizations here were made in \textsc{python} using \texttt{networkx.expected\_degree\_graph($\cdot$)}. From top to bottom the heatmaps represent the proportional L1 error of our estimate versus the actual output of \texttt{networkx.expected\_degree\_graph($\cdot$)}, the proportional L1 error between the input degree sequence versus \texttt{networkx.expected\_degree\_graph($\cdot$)}, and the ratio between the two proportional L1 errors respectively. We see that the proportional error between our estimate and actual Chung-Lu output is far smaller than that between the input and actual Chung-Lu output.}
    \label{fig:all_err}
\end{figure}

\subsection*{Predicting \naive Chung-Lu error}

\begin{figure}[!htb]
     \centering
     \includegraphics[width=0.42\textwidth]{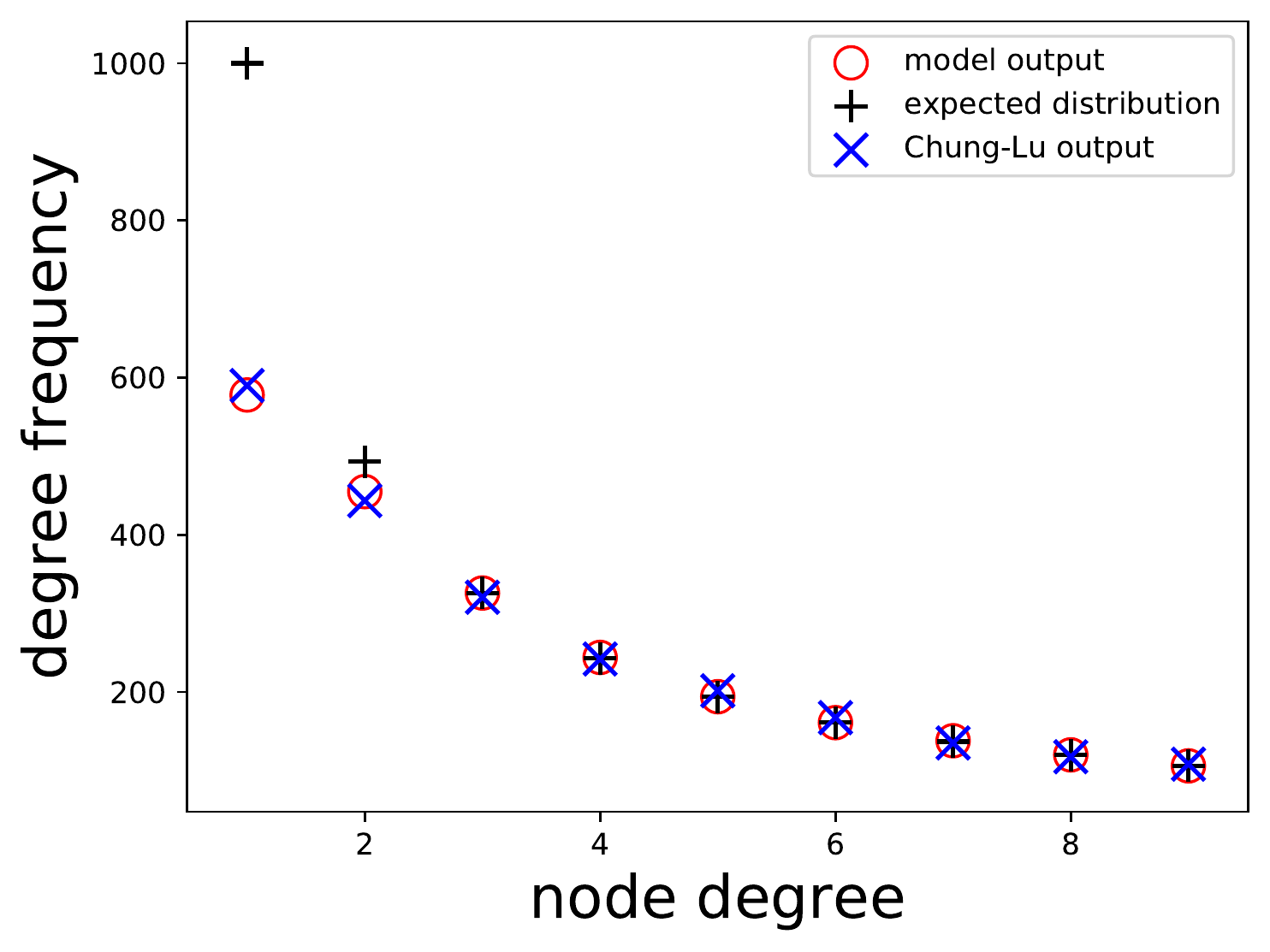}
     \includegraphics[width=0.42\textwidth]{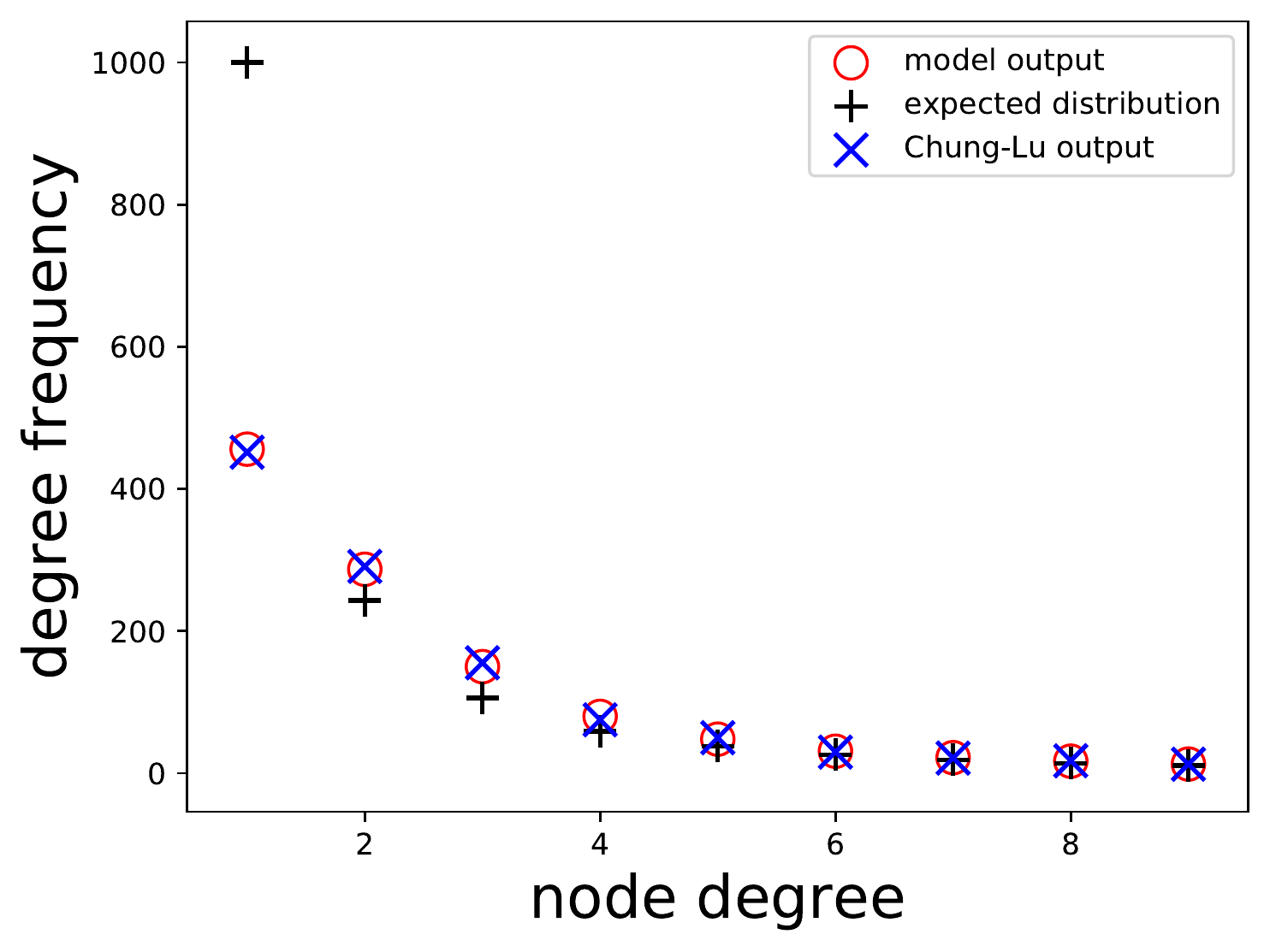}
      \includegraphics[width=0.42\textwidth]{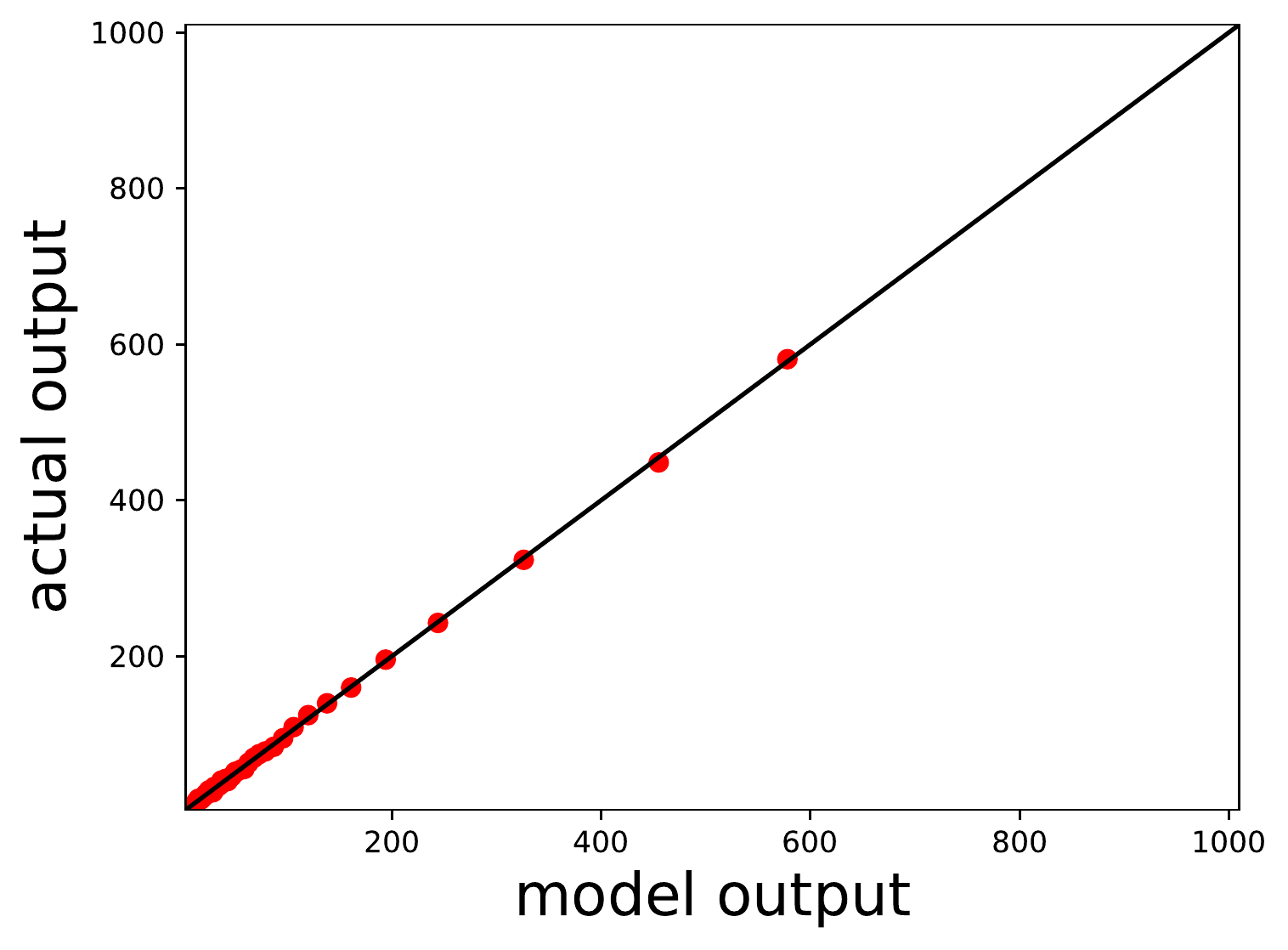}
     \includegraphics[width=0.42\textwidth]{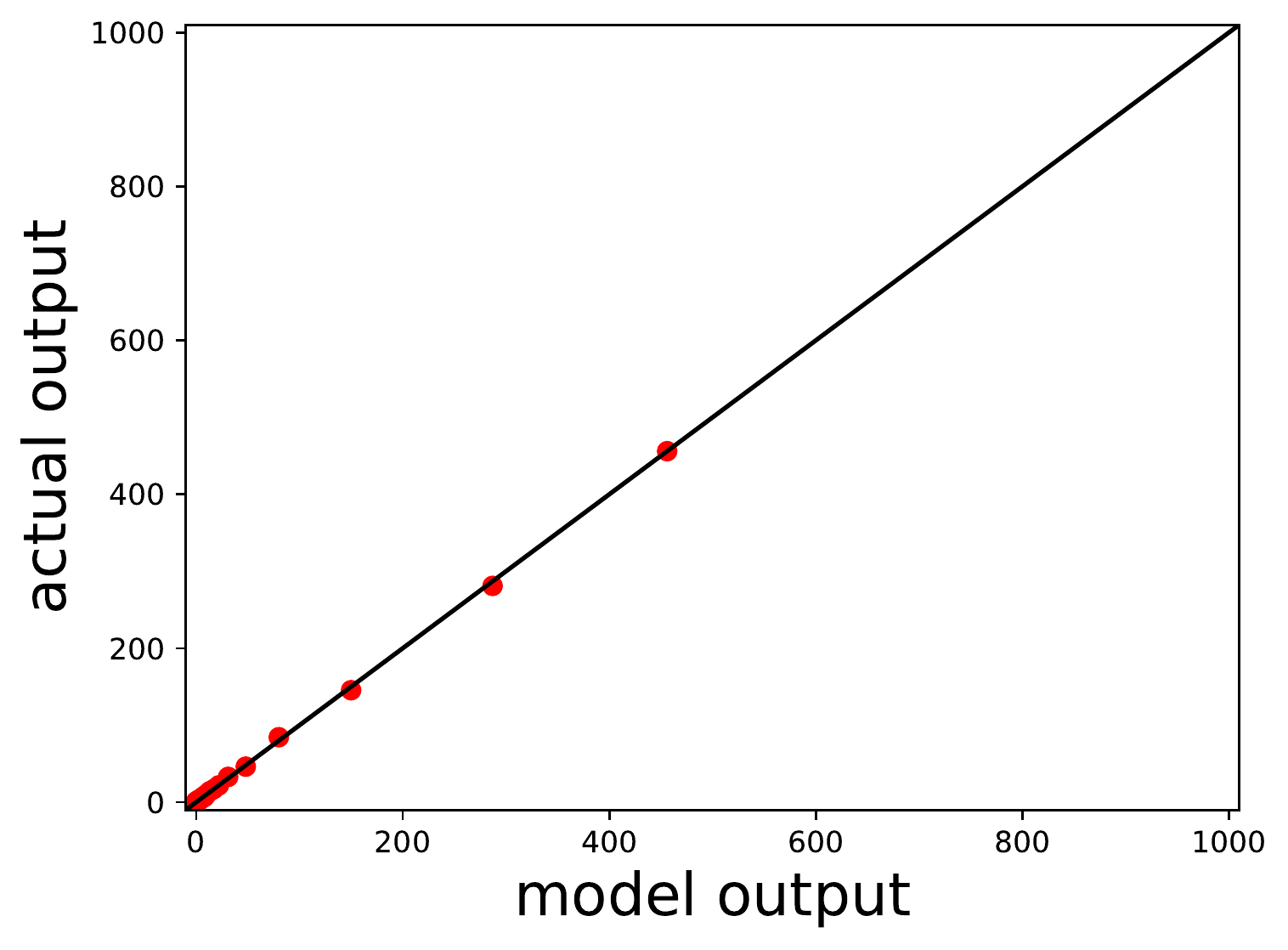}
     
     \caption{\textbf{Model distribution versus Chung-Lu outputs.} We consider the degree classes between one and nine for two different power law distributions. On the left is a power-law distribution with exponent $\beta = 1.0$ and on the right is a power-law distribution with exponent $\beta = 2.0$. In the top two plots, black crosses represent the \naive input $1000\times{k}^{-\beta}$, red circles represent the distribution our model estimates will be the output of Chung-Lu generation, and blue x's represent the average distribution for 20 instances of Chung-Lu graphs given the black crosses as input. We can see that the Chung-Lu generated graphs match our model output remarkably closely. The bottom two plots show how closely the Chung-Lu output distribution matches the input distribution up to the degree class 40 and we can see that all degree classes are close enough to touch the line denoting equivalence.}
     \label{fig:modeldiffplots}
 \end{figure}

We wish to determine how well $\textbf{P}$ models the output of the Chung-Lu algorithm for a given input distribution. We tested the approximation abilities of the matrix $\textbf{P}$ on a suite of 1950 power-law distribution inputs with $N_k = Nk^{-\beta}$ where $N$ was varied between $10^3$ and $10^5$ and $\beta$ was varied between 1 and 6 in intervals of 0.2. Additionally for these graphs we take all $m = 100$. The results of this can be seen in Figure~\ref{fig:all_err}. We find that while calculating the expected output with $\textbf{P}$ has near ten percent error for large $\beta$, it performs significantly better than {\naive}ly assuming the input distribution $x$ will resemble the output distribution $y$, which can result in proportional L1 errors near 80 percent. Unfortunately we cannot solve for the required input of general power-law distributions via $\textbf{P}^{-1}y$ since most power-law distributions fall outside of the valid, positive region of $\textbf{P}^{-1}y$ shown for $\mathbb{R}^4$ in Figure~\ref{fig:2dproj}. Additionally in Figure~\ref{fig:modeldiffplots} we compare the \naive output distribution to the outputs of both Chung-Lu generation and our model taking that distribution as input. We find that our model predicts the node degree frequency remarkably well.

\subsection*{Shifted Chung-Lu error}

\begin{figure}[!htb]
     \centering
     \includegraphics[width=0.7\textwidth]{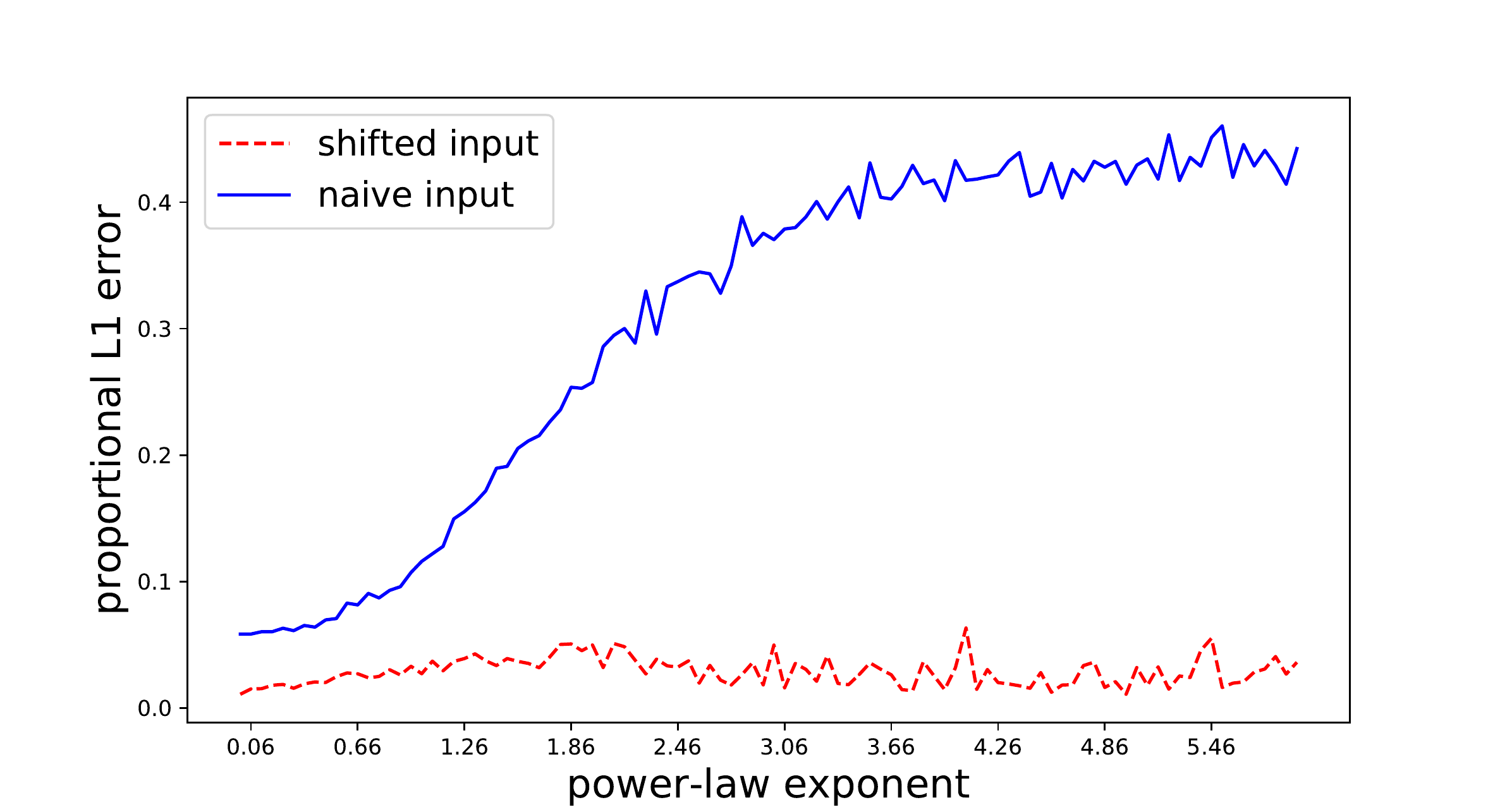}
     
     \caption{\textbf{Error of \naive Chung-Lu input versus shifted Chung-Lu input.} We consider 100 input distributions $Y_t$ such that $\textbf{P}^{-1}Y_t = X_t$ where the distribution $X_t$ is the power-law distribution $1000\times{k}^{-\frac{6t}{100}}$ with $k$ ranging between 1 and 40. For each of the 100 inputs, 30 graphs were generated and their degree distributions were averaged using the input $Y_t$ for Chung-Lu. The proportional L1 error between this output and the desired output $Y_t$ is shown as the solid blue line. Additionally 30 graphs were generated and their degree distributions were averaged using the input $X_t$ for Chung-Lu. The proportional L1 error between this output and the desired output $Y_t$ is shown as the dashed red line. We can see that the ``shifted" input we get using our model drastically reduces error for the sample. }
     \label{fig:shiftedinput}
 \end{figure}

We aim to determine how much proportional L1 accuracy is gained by using the vector $x=\textbf{P}^{-1}y$ as opposed to $y$ itself as an input to Chung-Lu. Since we have a closed form inverse of $\textbf{P}$ we should expect this accuracy to be comparable to the predictive accuracy of our model given in Figure~\ref{fig:all_err}, but this may not be the case numerically since $\textbf{P}^{-1}$ has elements below and above standard floating point precision. In the last subsection we noted that there is no positive inverse $\textbf{P}^{-1}y = x$ for many power law distributions. Due to this we generate a suite of 100 degree distributions of the form $y = [N_1,N_2,\cdots,N_m]^T$ with $m=40$. Here $N_k = 1000\times{k}^{-\beta}$ with $\beta$ varying between 0 and 6 in intervals of $6/100$ and we apply $\textbf{P}$ to these distributions to generate our test suite. These are guaranteed to be invertable distributions in the sense that $x{\in}\mathbb{R}^{+m}$ and $y{\in}\mathbb{R}^{+m}$. Then using our calculated $\textbf{P}^{-1}$ we compute and round the inputs yielding said outputs to the nearest integer vector. For this we use the variable precision toolbox in \textsc{matlab} with the digits of precision set to 100. The results of this can be seen in Figure~\ref{fig:shiftedinput}. We find that our shifted input drastically decreases the proportional L1 error between the output of Chung-Lu and the desired output. 

\section*{Conclusion}
We have provided a simple method for estimating the output of Chung-Lu random graph generators with far lower proportional L1 error than that given by the traditional assumption that output distributions will resemble input distributions. Our method utilized a Poisson estimate for the number of nodes of given degrees and we used this to define an invertible matrix $\textbf{P}$ that models the expected output from Chung-Lu generators. This allowed us to ``solve the problem in reverse'' and take a desired output $y$ and solve for the Chung-Lu input $x$ that will result in $y$. We called this the shifted Chung-Lu input. We showed $\textbf{P}$ predicts that many degree distributions simply are not feasible for Chung-Lu generators, however we provide two conditions for determining when a desired output is feasible. We showed that this method, while mathematically simple, has some numerical drawbacks and requires an excessive amount of numerical precision for accuracy. However, the $\textbf{P}$ matrix can be reused and this work provides significant accuracy increases with trivial algorithmic cost once the $\textbf{P}$ matrix is calculated and stored. 

\indent There are several avenues for further research. For instance, this work lends itself to analysis and improvement of graph generation. Methods which use \naive Chung-Lu generation as a subroutine may gain both accuracy and insight into possible distribution errors through the kind of analysis done in this paper. Further work may also be done on how altering connection probabilities between degree classes may be used to fine tune the matrix $\textbf{P}$ in order to produce graphs which are inadvisable for \naive Chung-Lu generation.

\bibliographystyle{siamplain}
\clearpage
\bibliography{reference}

\end{document}